\documentclass{PoS}
\usepackage{amsmath}
\usepackage[nosort]{cite}

\newcommand{\beq}{\begin{equation}}
\newcommand{\eeq}{\end{equation}}
\newcommand{\be}{\begin{equation}}
\newcommand{\ee}{\end{equation}}
\newcommand{\bea}{\begin{eqnarray}}
\newcommand{\eea}{\end{eqnarray}}
\newcommand{\ben}{\begin{eqnarray*}}
\newcommand{\een}{\end{eqnarray*}}               
\newcommand{\ba}{\begin{aligned}}
\newcommand{\ea}{\end{aligned}}
\newcommand{\bt}{\begin{tabular}}
\newcommand{\et}{\end{tabular}}
\newcommand{\bc}{\begin{center}}
\newcommand{\ec}{\end{center}}

\newcommand{\cS}{\mathcal{S}}
\newcommand{\cK}{\mathcal{K}}
\newcommand{\cN}{\mathcal{N}}

\newcommand{\cV}{\mathcal{V}}

\newcommand{\cref}{{\bf [check ref]}}

\title{On Mirror Symmetry for Calabi-Yau Fourfolds with Three-Form Cohomology}

\ShortTitle{On Mirror Symmetry for Calabi-Yau Fourfolds with Three-Form Cohomology}

\author{\speaker{Sebastian Greiner}\\

Max-Planck-Institut f\"ur Physik, \\
F\"ohringer Ring 6, 80805 Munich, Germany \\
and \\
Institute for Theoretical Physics and \\
Center for Extreme Matter and Emergent Phenomena, \\
Utrecht University, Leuvenlaan 4, 3584 CE Utrecht, The Netherlands \\
\\
E-mail: \email{sgreiner@mpp.mpg.de}}

\abstract{We review the Kaluza-Klein reduction of Type IIA string theory on Calabi-Yau fourfolds and apply mirror symmetry to the resulting two-dimensional $ \mathcal{N}=(2,2) $ effective theories. In the course of the reduction we focus especially on non-trivial three-form cohomology on these fourfolds and investigate the couplings of the corresponding massless zero-modes. These show a dependence on both complex structure as well as K\"ahler structure deformations and we provide evidence that they are determined by two holomorphic functions that get exchanged via mirror symmetry. Application of the mirror map enables us to give an explicit description of these functions at the large volume and large complex structure point of the moduli space.}

\FullConference{Corfu Summer Institute 2016 "School and Workshops on Elementary Particle Physics and Gravity"\\
31 August - 23 September, 2016\\
Corfu, Greece}

\begin{document}

\section{Introduction}

\noindent
Low-energy effective actions in four spacetime dimensions can be obtained from string theory by a Kaluza-Klein reduction on a compact internal geometry. In the context of F-theory \cite{Vafa:1996xn,Denef:2008wq,Weigand:2010wm} this internal geometry is described by an elliptically fibered Calabi-Yau fourfold \cite{Greene:1993vm,Mayr:1996sh,Klemm:1996ts,Braun:2014xka} and the most convenient way to derive its effective low-energy action is via the M- to F-theory limit described in \cite{Grimm:2010ks}. In this review we focus on the expansion for the massless zero-modes of harmonic three-forms into $ (2,1) $-forms proposed in \cite{Grimm:2010ks} and the resulting scalar fields in the effective theory. These have couplings that depend on the complex structure as well as the K\"ahler moduli, which was found in reductions of M-theory in \cite{Haack:2000di}, in reductions of type IIA in \cite{Haack:2001jz} and for F-theory in \cite{Grimm:2010ks}.

The goal of our original work with Thomas Grimm \cite{Greiner:2015mdm} is to establish an ansatz for a general fourfold reduction and to calculate the complex structure dependence of the harmonic $ (2,1) $-forms. A particular simple example of a dimensional reduction on a Calabi-Yau fourfold background is type IIA supergravity with a Calabi-Yau fourfold as internal space  and a two-dimensional Minkowski spacetime as external space as discussed in \cite{Gates:2000fj,Haack:2001jz,Gukov:2002iq}. This is a convenient toy-example, since the theory contains a three-form gauge-field and a two-dimensional effective theory has no propagating vector degrees of freedom. In addition, the effective two-dimensional supergravity theory we obtain has a very simple mirror map, i.e.~it has a dual description in terms of a different Calabi-Yau fourfold background. This allows a computation of the complex structure dependence of the harmonic three-forms in which we expand the three-form gauge-potential.

This article is structured as follows. In section \ref{geometry-section} we establish the basics of the geometry and topology of a Calabi-Yau fourfold and introduce the expansion of a harmonic three-form into harmonic $ (2,1) $-forms. We use this in section \ref{reduction-section} to review the dimensional reduction of type IIA supergravity on a Calabi-Yau fourfold geometry. We also discuss the $ \cN=(2,2) $ dilaton-supergravity effective action that we obtain as a result of the reduction and its dual formulation where we change the representation of the three-form moduli from chiral to twisted-chiral multiplets. The effective theory allows for a simple mirror map that we review in section \ref{mirror-section}. Here we also derive the key result of this work, the complex structure dependence of our three-form ansatz at the large volume/large complex structure point of the moduli space of the effective theory. We conclude with a summary and give a short overview of further progress in the field in section \ref{summary-section}.

\section{On the geometry of Calabi-Yau fourfolds with three-form cohomology} \label{geometry-section}

\noindent
In this section we review the geometrical and topological properties of a general Calabi-Yau fourfold. In addition, we highlight the special ansatz we make for the three-form cohomology on the fourfold. This facilitates the introduction of a matrix valued function $ f_{mn} $ holomorphic in the complex structure moduli that will be of key importance throughout this work.

For our considerations, a Calabi-Yau fourfold $ Y_4 $ is a compact complex manifold of eight real dimensions with holonomy group the full $ SU(4) $. This implies already that the manifold is K\"ahler and enables us to find a closed two-form $ J $, the K\"ahler form and a real function, the K\"ahler potential, that defines the metric. Calabi-Yau manifolds have their physical relevance because they allow for a unique Ricci-flat metric within the class of $ J $. Due to the holonomy group being all of $ SU(4) $, we can find a complex covariantly constant and no-where vanishing spinor of definite chirality and we can obtain $ J $ as a bilinear contraction using this spinor. One of the possible bilinear contractions of this spinor provides the unique non-trivial holomorphic $ (4,0) $-form $ \Omega $ that defines a trivial line-bundle over the fourfold. For the sake of completeness, we also introduce the total volume $ \mathcal{V} $ of $ Y_4 $ that is related to $ \Omega $ and its complex conjugate as follows
\begin{equation}
 \mathcal{V} = \frac{1}{4!} \int_{Y_4} J \wedge J \wedge J \wedge J \, , \quad |\Omega|^2 = \frac{1}{\mathcal{V}^2} \int_{Y_4} \Omega \wedge \bar \Omega \, .
\end{equation}
On a K\"ahler manifold we also have a Hodge decomposition on the cohomology groups, the dimensions of the corresponding groups are called Hodge numbers and are defined to be $h^{p,q}(Y_4) = \mathrm{dim}_{\mathbb{C}}(H^q(Y_4, \Omega^p)) $. These are the eigenspaces of the induced autormorphism on $ H^{p+q}(Y_4, \mathbb{C}) $ of the complex structure of $ Y_4 $. The three independent Hodge numbers are $ h^{1,1}(Y_4), h^{3,1}(Y_4) $ and $ h^{2,1}(Y_4) $. Their physical interpretation is the massless field content in the effective theory of the dimensional reduction performed in the upcoming section. Similar to the threefold case, $ h^{1,1}(Y_4) $ counts the possible deformations of the K\"ahler structure, denoted by $ v^A $ and $ h^{3,1} $ counts the number of complex structure deformations of $ Y_4 $, denoted by $ z^K $, both preserving the Calabi-Yau condition. These correspond to moduli fields in the effective theory obtained by Kaluza-Klein reduction on $ Y_4 $. In contrast to threefolds, Calabi-Yau fourfolds may also have an additional independent sector of massless modes, that do not have a simple geometrical interpretation. These modes arise from cohomological non-trivial three-forms, counted by $ h^{2,1}(Y_4) $, and the purpose of this work is to investigate the physical consequences of a non-trivial three-form cohomology.\\
The important tool for our considerations will be mirror symmetry. This physical duality has its geometrical manifestation in its most simple form by the postulate, that for each $ Y_4 $ there exists a second Calabi-Yau fourfold $ \hat Y_4 $ with Hodge numbers flipped along the diagonal of the Hodge diamond indicated in the following graphic displaying the non-trivial Hodge numbers of a Calabi-Yau fourfold.\\

\vspace{.5cm}
\arraycolsep=2,0pt\def\arraystretch{1.4}
\setlength{\unitlength}{0.6cm}
\begin{picture}(21,8)
 \put(5,8){$ h^{0,0} $}
 \put(4,7){$ h^{1,0} $} \put(6,7){$ h^{0,1} $}
 \put(3,6){$ h^{2,0} $} \put(5,6){$ h^{1,1} $} \put(7,6){$ h^{0,2} $}
 \put(2,5){$ h^{3,0} $} \put(4,5){$ h^{2,1} $} \put(6,5){$ h^{1,2} $} \put(8,5){$ h^{0,3} $}
 \put(1,4){$ h^{4,0} $} \put(3,4){$ h^{3,1} $} \put(5,4){$ h^{2,2} $} \put(7,4){$ h^{1,3} $} \put(9,4){$ h^{0,4} $}
 \put(2,3){$ h^{4,1} $} \put(4,3){$ h^{3,2} $} \put(6,3){$ h^{2,3} $} \put(8,3){$ h^{1,4} $}
 \put(3,2){$ h^{4,2} $} \put(5,2){$ h^{3,3} $} \put(7,2){$ h^{2,4} $}
 \put(4,1){$ h^{4,3} $} \put(6,1){$ h^{3,4} $}
 \put(5,0){$ h^{4,4} $}
 
 \put(11,4){$ = $}
 
 \put(17,8){$ 1 $}
 \put(16,7){$ 0 $} \put(18,7){$ 0 $}
 \put(15,6){$ 0 $} \put(17,6){$ h^{1,1} $} \put(19,6){$ 0 $}
 \put(14,5){$ 0 $} \put(16,5){$ h^{2,1} $} \put(18,5){$ h^{2,1} $} \put(20,5){$ 0 $}
 \put(13,4){$ 1 $} \put(15,4){$ h^{3,1} $} \put(17,4){$ h^{2,2} $} \put(19,4){$ h^{3,1} $} \put(21,4){$ 1 $}
 \put(14,3){$ 0 $} \put(16,3){$ h^{2,1} $} \put(18,3){$ h^{2,1} $} \put(20,3){$ 0 $}
 \put(15,2){$ 0 $} \put(17,2){$ h^{1,1} $} \put(19,2){$ 0 $}
 \put(16,1){$ 0 $} \put(18,1){$ 0 $}
 \put(17,0){$ 1 $}
 \setlength{\unitlength}{0.06cm}
 \multiput(20,73)(10,-10){7}{\line(1,-1){6}}       
\end{picture}
\\

\noindent 
A more elaborate discussion of the geometry of Calabi-Yau fourfolds can be found in \cite{Mayr:1996sh,Klemm:1996ts}.

Let us now turn our attention towards the non-trivial three-forms. Due to $ h^{3,0}(Y_4) = 0 $, we can find a basis of the three-form cohomology $ H^3(Y_4, \mathbb{C}) $ comprised by $ (2,1) $-forms and their complex conjugates. As argued in \cite{Greiner:2017ery}, such a basis can be chosen to be of the form
\begin{equation}
 \psi_l = \alpha_l + i \, f_{lm}(z) \, \beta^m \quad \in \, H^{2,1}(Y_4) \, , \quad l,m = 1,\ldots,h^{2,1}(Y_4) \, ,
\end{equation}
where $ (\alpha_l, \beta^m) $ provide a basis for the real three-forms $ H^3(Y_4,\mathbb{R}) $ and are hence topological.
\footnote{The difference between $ \alpha_l $ and $ \beta^m $ is that we choose $ \beta^l \wedge \beta^m = 0 $ which is the case for Calabi-Yau fourfolds realized as toric hypersurfaces described in \cite{Greiner:2017ery}. The vanishing of this intersection is, however, not necessary as was found in \cite{Corvilain:2016kwe}.}
The whole complex structure dependence of the basis is encoded in the holomorphic matrix valued function $ f_{lm}(z) $ that depend a priori on the full set of complex structure moduli $ z^K $. Such a basis is in particular holomorphic
\begin{equation}
 \frac{\partial}{\partial \bar z^K} \, \psi_l = 0
\end{equation}
for all $ l= 1, \ldots, h^{2,1}(Y_4) $ and $ K = 1, \ldots, h^{3,1}(Y_4) $. This base choice has the additional property that the matrix $ \mathrm{Re}(f)_{lm} $ is invertible and hence we can normalize this basis with the inverse $ \mathrm{Re}(f)^{lm} $ of $ \mathrm{Re}(f)_{lm} $ to the more convenient basis
\begin{equation} \label{Psi_def}
 \Psi^l = \frac{1}{2} \, \mathrm{Re}(f)^{lm}(\alpha_m - i \, \bar f_{mn}(\bar z) \, \beta^n) \quad \in \, H^{1,2}(Y_4) \,  ,
\end{equation}
that has the advantage that $ \partial_K \Psi^l \in H^{1,2}(Y_4) $. The reason for demanding a basis of three-forms with a definite type, is the action of the Hodge star operator $ \ast $ on them
\begin{equation}
 \ast \, \psi =  i \, J \wedge \psi \, , \quad \ast \, \bar \psi = -i \, J \wedge \psi \, , \quad \psi \in H^{2,1}(Y_4) \, ,
\end{equation}
where $ J $ is the K\"ahler form of $ Y_4 $. The introduced basis $ \Psi^l $ will be useful for the dimensional reduction of type IIA supergravity on $ Y_4 $ to which we will turn next.

\section{Dimensional reduction of Type IIA supergravity} \label{reduction-section}

\noindent
In this section we review the Kaluza-Klein reduction of the low-energy limit of type IIA string theory, type IIA supergravity, on a general Calabi-Yau fourfold geometry times a two-dimensional Minkowski spacetime. Assuming the submanifolds of the internal Calabi-Yau space to have volume much larger than the string scale, we can justify to consider only the massless zero-modes in the low-energy limit. Similar reductions were already performed in \cite{Haack:2000di,Gates:2000fj,Gukov:2002iq}.

Starting point of our reduction procedure is the bosonic part of the ten-dimensional action of type IIA supergravity in string frame given by
\begin{align}
 S^{(10)} _{IIA} &= \int e^{-2 \check \phi_{IIA}} \Big( \, \frac{1}{2} \,  \check R \, \check \ast \,  \check 1 + 2 \, d \check \phi_{IIA} \wedge \check \ast \,  d \check \phi_{IIA} - \frac{1}{4} \, \check H _3 \wedge \check \ast \,  \check H_3 \, \Big) \nonumber \\
 &\quad - \frac{1}{4} \int \Big( \, \check F_2 \wedge \check \ast \, \check F_2 + \check{\textbf{F}}_4 \wedge \check \ast \, \check{\textbf{F}}_4 + \check B_2 \wedge \check F_4 \wedge \check F_4 \, \Big) \, .
\end{align}
The field content aside of the regular Einstein-Hilbert term is as follows. We have the ten-dimensional dilaton $ \check \phi_{IIA} $, the field strength $ \check H_3 = d \check B_2 $ of the NS-NS two-form $ \check B_2 $ and $ \check F_{p+1} = d \check C_p $ are the field strengths of the R-R $ p $-forms $ \check C_1 $ and $ \check C_3 $. It is convenient to introduce the modified field strength $ \check{\textbf{F}}_4 = \check F_4 - \check C_1 \wedge \check H_3 $. The check above symbols will indicate here and in the following ten-dimensional fields and operators.

The background around which we will expand the fields will be of the form $ \mathbb{M}_{1,1} \times Y_4 $ where $ \mathbb{M}_{1,1} $ is the two-dimensional extended Minkowski spacetime and $ Y_4 $ the Calabi-Yau fourfold which provides the compact internal space. Since $ Y_4 $ has a covariantly constant spinor of definite chirality, the dimensional reduction of the $ \cN=(1,1) $ supersymmetry in ten dimensions will lead to $ \cN=(2,2) $ supersymmetry in the two-dimensional effective theory. Note that we do not enable fluxes, i.e.~the $ p $-form field strengths have zero vacuum expectation value and we choose the unique Ricci-flat metric on $ Y_4 $ as a background metric.

Let us go throught the massless perturbations around this background that will preserve supersymmetry, i.e.~the Calabi-Yau condition on $ Y_4 $. These perturbations are comprised of the fluctuations of the internal K\"ahler metric $ g_{i\bar j} $, the fluctuations of the $ p $-form fields $ \check B_2, \check C_1, \check C_3 $ and the dilaton $ \check \phi_{IIA} $. The perturbations will provide the bosonic degrees of freedom of the $ \cN=(2,2) $ scalar multiplets, since there are no propagating vectors in two spacetime dimensions.

The K\"ahler structure deformations are described by real moduli $ v^A $, $ A= 1, \ldots, h^{1,1}(Y_4) $ preserving the complex structure of $ Y_4 $ and are given by
\begin{equation}
 g_{i \bar j} + \delta g_{i \bar j} = -i \, J_{i \bar j} = -i \, v^A (\omega_A)_{i \bar j}
\end{equation}
where $ J $ is the K\"ahler form of $ Y_4 $ and $ \omega_A $ is a real basis of $ H^{1,1}(Y_4) $. These combine with the real moduli $ b^A $ of the expansion of the $ \check B_2 $ field around its zero vacuum expectation value
\begin{equation}
 \check B_2 = b^A \, \omega_A
\end{equation}
into the bosonic part of twisted-chiral multiplets
\begin{equation}
 t^A = b^A + i \, v^A
\end{equation}
of the resutling $ \cN =(2,2) $ supergravity. The other fluctuations around the vacuum metric leaving the K\"ahler structure of $ Y_4 $ invariant are parametrized by the complex structure moduli $ z^K $, $ K= 1, \ldots, h^{3,1}(Y_4) $ and given by
\begin{equation}
 \delta g_{\bar i \bar j} = \frac{1}{3|\Omega|^2} \, \bar \Omega_{\bar i} {}^{lmn} (\chi_K)_{lmn \bar j} \, \delta z^K
\end{equation}
for $ \chi_K $ a basis of $ H^{3,1}(Y_4) $ and $ \Omega $ the holomorphic $ (4,0) $-form of $ Y_4 $. These moduli $ z^K $ reside in the bosonic part of chiral multiplets in the effective two-dimensional theory.

Since there are no propagating vectors in two dimensions and $ h^{1,0}(Y_4) = 0 $, we do not need to consider $ \check C_1 $ in our discussion and the decomposition of $ \check C_3 $ provides only additional complex scalar moduli $ N^l $, $ l=1,\ldots, h^{2,1}(Y_4) $. \footnote{This is possible, since we do not obtain any gaugings in the effective theory.} Here we use the ansatz introduced in \eqref{Psi_def} to expand $ \check C_3 $ into three-forms of the internal space
\begin{equation}
 \check C_3 = N_l \, \Psi^l + \bar N_l \, \bar \Psi^l \, .
\end{equation}
The complex scalars $ N_l $ also become the bosonic part of chiral multiplets of the $ \cN=(2,2) $ supergravity theory. To conclude our discussion of massless fluctuations, we also expand the ten-dimensional dilaton by dropping its dependence on the internal space and shifting it with the (logarithm of the) volume $ \cV $ of $ Y_4 $ to obtain
\begin{equation}
 e^{2\phi_{IIA}} = \frac{e^{2\check \phi_{IIA}}}{\cV} \, , \quad \cV = \frac{1}{4!} \int_{Y_4} J^4 \, .
\end{equation}
The real dilaton will be part of the gravity multiplet of the resulting two-dimensional $ \cN=(2,2) $ supergravity, which is hence also known as $ \cN=(2,2) $ dilaton-supergravity. The corresponding general bosonic action of this theory is
\begin{align} \label{N2_dil_action}
 S^{2} _{dil} = \int e^{-2\tilde \varphi} & \Big( \, \frac{1}{2} \, R \ast 1 + 2 d \tilde \varphi \wedge \ast \, d\tilde \varphi 
 - \cK_{\phi^\kappa \bar \phi^\lambda} d \phi^\kappa \wedge \ast \, d \bar \phi^\lambda 
 + \cK_{\sigma^A \bar \sigma^B} d \sigma^A \wedge \ast \, d \bar \sigma^B \nonumber \\
 &\quad - \cK_{\phi^\kappa \bar \sigma^B} d \phi^\kappa \wedge d \bar \sigma^B
 - \cK_{\sigma^A \bar \phi^\lambda} d \bar \phi^\lambda \wedge d \sigma^A \, \Big) \, ,
\end{align}
where $ \cK $ is a real function of the chiral multiplets $ \phi^\kappa $ and the twisted-chiral multiplets $ \sigma^A $. We denoted derivatives of $ \cK $ with respect to the fields by $ \cK_{\phi^\kappa \bar \phi^\lambda} = \partial_{\phi^\kappa} \partial_{\bar \phi^\lambda} \cK $ and similar for the remaining fields. For the Kaluza-Klein reduction on the Calabi-Yau fourfold $ Y_4 $ with $ h^{2,1} = 0 $ the perturbations can be identified with the fields appearing in the general action as 
\begin{equation}
 \tilde \varphi = \phi_{IIA} \, , \quad \phi^\kappa = z^K \, , \quad \sigma^A = t^A \, . 
\end{equation}
The generating function $ \cK $ of the coefficient matrices is found to be
\begin{equation} \label{cK}
 \cK = - \, \mathrm{log} \int_{Y_4} \Omega \wedge \bar \Omega \, + \, \mathrm{log} \, \cV \, .
\end{equation}
This was already established in \cite{Gates:2000fj} where also the properties of the $ \cN=(2,2) $ dilaton-supergravity are discussed in detail using the superspace formalism. The extension of this theory we proposed in \cite{Greiner:2015mdm} was to include the three-form moduli $ N_l $ and modify $ \cK $ to obtain the field identifications
\begin{equation}
 \tilde \varphi = \phi_{IIA} \, , \quad \phi^\kappa = (z^K, N_l) \, , \quad \sigma^A = t^A \, . 
\end{equation}
and the generating function
\begin{equation} \label{tilde_cK}
 \tilde \cK = \cK + e^{-2 \tilde \varphi} \cS \, , \quad \cS = \mathrm{Re}(N)_l \, \mathrm{Re}(N)_m \, H^{lm} \, .
\end{equation}
Here we introduced $ H^{lm} $ which is the metric on the three-form moduli space defined by
\begin{equation}
 H^{lm} = \int_{Y_4} \Psi^l \wedge \ast \, \bar \Psi^m = i \int_{Y_4} J \wedge \Psi^l \wedge \bar \Psi^m \, .
\end{equation}
Using the ansatz of \eqref{Psi_def}, we can split this metric into
\begin{equation} \label{H_ref}
 H^{lm} = -\frac{1}{2} \, \mathrm{Re}( f)^{ln} \, \mathrm{Re}(h)_n {}^m
\end{equation}
where $ f_{ln} (z) $ is the function holomorphic capturing the dependence on complex structure moduli $ z^K $ of the three-form basis $ \Psi^l $ and 
\begin{equation} \label{hC_def}
 h_n {}^m = -i \, t^A \, C_{An} {}^m \, , \quad C_{An} {}^m = \int_{Y_4} \omega_A \wedge \alpha_n \wedge \beta^m \, ,                                                                                                                           
\end{equation}
is a holomorphic function of the complexified K\"ahler moduli $ t^A $ and $ C_{An} {}^m $ is a real and topological intersection number constant over the moduli space.

Due to the shift-symmetry $ N_l \rightarrow N_l + i c_l $ of the effective theory for $ c_l $ a real vector, we can find a dual $ \cN=(2,2) $ dilaton-supergravity where the three-form moduli degrees of freedom are represented by twisted-chiral multiplets with bosonic part the complex scalars $ {N^\prime}^l $ and the rest of the field content preserved. The precise relations are given by
\begin{align}
 {N^\prime}^l = \frac{1}{2} \, \frac{\partial \cS}{\partial \mathrm{Re} (N)_l} + i {\lambda^\prime} ^l = H^{lm} \, \mathrm{Re} (N)_m + i \, {\lambda^\prime}^l
\end{align}
with the real scalars $ {\lambda^\prime}^l $ dual to $ N_l - \bar N_l $. The new kinetic function $ \tilde \cK^\prime $ is determined to be
\begin{equation} \label{tilde_cK_prime}
 \tilde \cK^\prime = \cK - e^{2\phi_{IIA}} H_{kl} \, \mathrm{Re} (N^\prime)^k \, \mathrm{Re} (N^\prime)^l
\end{equation}
where $ H_{kl} $ is the inverse of the metric on the three-form moduli introduced in \eqref{H_ref}. For more details we refer to the original work in \cite{Greiner:2015mdm}. This different representation of the same theory with the degrees of freedom corresponding to three-form moduli residing in chiral or twisted-chiral multiplets will be useful when we apply mirror symmetry to our theory to which we will come in the following section.

\section{Mirror symmetry at large volume and large complex structure} \label{mirror-section}

\noindent
In this section we want to apply mirror symmetry to our effective theory. Mirror symmetry states in our context that for type IIA string theory compactified on a Calabi-Yau fourfold $ Y_4 $ there exists a mirror Calabi-Yau fourfold $ \hat Y_4 $ such that type IIA string theory on $ \hat Y_4 $ leads to the same physics as the compactification on $ Y_4 $. This holds in particular for the low-energy effective theory at large volume we considered in the previous section. In the following we will introduce the geometric analogs of quantities on $ \hat Y_4 $ with hatted symbols. For details of the mirror symmetry discussion we refer to the original work \cite{Greiner:2015mdm} and \cite{Gukov:2002iq}.

As was found in \cite{Gukov:2002iq} mirror symmetry leaves the pure $ \cN=(2,2) $ dilaton-supergravity we obtained from the dimensional reduction invariant. Its only consequence is an exchange of chiral and twisted-chiral multiplets preserving the structure of the action as in \eqref{N2_dil_action}. At the large complex structure and large volume point in moduli space, this implies that the chiral multiplets $ z^K $ describing complex structure variations of $ Y_4 $ map to the twisted-chiral multiplets $ \hat t^K $ parametrizing the K\"ahler structure of $ \hat Y_4 $ and vice versa $ t^A $ map to $ \hat z^A $. This is familiar from mirror symmetry of Calabi-Yau threefolds as described in \cite{Hosono:1993qy} and was generalized to higher dimensions in \cite{Greene:1993vm}. To obtain the same physical theory, the corresponding terms of the kinetic potential \eqref{cK} also get exchanged under mirror symmetry and change their sign, since the kinetic terms of chiral and twisted-chiral multiplets appear with opposite signs in the action \eqref{N2_dil_action}.

Since the three-form moduli are also represented by chiral multiplets $ N_l $ in the dimensional reduction on $ Y_4 $, mirror symmetry maps them to twisted chiral-multiplets $ \hat{ N^\prime}^l $ of the mirror theory on $ \hat Y_4 $. Correspondingly the kinetic potentials $ \tilde \cK $ of \eqref{tilde_cK} and $ \hat {\tilde \cK}^\prime $ derived in \eqref{tilde_cK_prime} of the mirror need to be matched. This facilitated the dualization procedure we eluded on in the previous section \ref{reduction-section}. The matching of the two mirror theories is summarized in the following list:
\begin{align}
 \big(\, t^A , \, N_l \, , \, z^K \, \big) \quad &\leftrightarrow \quad \big( \, \hat z^A , \, \hat {N^\prime}^l , \, \hat t^K \, \big)  \\
 \mathrm{log} \, \cV \quad &\leftrightarrow \quad - \,\mathrm{log} \int_{\hat Y_4} \hat \Omega \wedge \hat {\bar \Omega} \\
 \mathrm{log} \int_{Y_4} \Omega \wedge \bar \Omega \quad &\leftrightarrow \quad -\, \mathrm{log} \, \hat V \\
 H^{lm} \quad &\leftrightarrow \quad \hat H_{lm} \, .
\end{align}
Note that we found that the metric $ H^{lm} $ on the three-form moduli space gets mapped to the inverse $ \hat H_{lm} $ of its counterpart $ \hat H^{lm} $ on the mirror geometry. Recall that we know the K\"ahler moduli dependence given by $ h^n {}_m (t^A) $ of $ H^{lm} $ in detail
\eqref{H_ref}
\begin{equation}
 H^{lm} = -\frac{1}{2} \, \mathrm{Re} (f)^{ln} \, \mathrm{Re} (h)^n {}_m \, , \quad h^n {}_m = -i \, t^A \, C_{Am} {}^n \, .
\end{equation}
Due to the matching of the mirror theories we can hence infer that the holomorphic matrix valued functions $ f_{lm} $ and $ \hat h^n {}_m $ of the mirror geometries can be identified
\begin{equation} \label{f_linear}
 f_{lm} (z^K) = -i \, z^K \, \hat C_{Kl} {}^m \, , \quad  \hat C_{Kl} {}^m = \int_{\hat Y_4} \hat \omega_K \wedge \hat \alpha_l \wedge \hat \beta^m \, ,
\end{equation}
where we introduced the intersection numbers $ \hat C_{Kl} {}^m $ as the mirror analog of \eqref{hC_def}. This identification should be viewed as valid around the large complex structure and large volume point of the underlying Calabi-Yau fourfold moduli space and is our main result. This concludes our discussion on mirror symmetry of Calabi-Yau fourfolds with non-trivial three-form cohomology at the complex structure and large volume point in moduli space.

\section{Summary and Outlook} \label{summary-section}

\noindent
In this article we reviewed the dimensional reduction of type IIA supergravity on a general Calabi-Yau fourfold keeping only the massless zero-modes. As a result we found a two-dimensional $ \cN=(2,2) $ dilaton-supergravity and determined the field content as well as the kinetic potential. We were able to include the massless zero-modes arising from three-form moduli by a special ansatz. In the course of the discussion we determined the right metric on the three-form moduli space. The shift-symmetry in the imaginary part of the three-form moduli allowed us to perform a dualization of the three-form moduli degrees of freedom from a representation as chiral multiplets into a representation as twisted-chiral multiplets. Invariance of the pure $ \cN=(2,2) $ dilaton-supergravity under mirror symmetry made a simple description of the mirror map possible. This map matches chiral with twisted-chiral multiplets of the mirror theories. At the large complex structure/large volume point we were able to compute the the metric on the three-form moduli space explicitly. The key insight we gained is that the matrix-valued holomorphic function $ f_{lm}(z) $ is linear in the complex structure moduli for large complex structure with leading coefficient an intersection number of the mirror geometry as shown in \eqref{f_linear}.

Our original work \cite{Greiner:2015mdm} also discusses first applications to M- and F-theory effective actions on Calabi-Yau fourfolds. There we write down the corresponding three- and four-dimensional effective actions and illuminate the connection to the two-dimensional theory we reviewed here, where the correct rescaling of the K\"ahler moduli with the dilaton plays an important role. We also comment on the weak coupling limit of F-theory \cite{Sen:1996vd,Sen:1997gv} and compare our results with the existing type IIB orientifold effective theory \cite{Grimm:2004ua} and find full agreement.

The shift symmetries of the three-form moduli are further discussed in \cite{Corvilain:2016kwe} where it is also explained that the three-form moduli space is a higher dimensional torus due to the shift symmetries of the three-form gauge potential. Explicit examples of Calabi-Yau fourfolds with non-trivial three-forms and their moduli space we construct in \cite{Greiner:2017ery} as toric hypersurfaces. In this context the $ (2,1) $-forms arise from holomorphic one-forms on Riemann surfaces within the Calabi-Yau fourfold and we give an interpretation of the function $ f_{lm} $ in terms of the periods of Riemann surfaces. This allows us to find differential equations, so called Picard-Fuchs equations, for $ f_{lm} $ for which the result \eqref{f_linear} provides the boundary conditions in moduli space. In addition, we realize in \cite{Greiner:2017ery} explicit examples of F-theory compactifications and their weak coupling limits.

Currently we are working on the construction of Calabi-Yau fourfolds as complete intersections in toric ambient spaces, a better understanding of the weak coupling limit of F-theory effective theories and first phenomenological applications of our new insights. Also a precise understanding of the calculation and interpretation of the geometrical mirror map is desirable and under discussion. Furthermore, we hope to shed light on the inclusion of fluxes and their relation to three-form moduli on elliptically fibered Calabi-Yau fourfolds with our explicit constructions. Since we found a procedure to calculate holomorphic coupling functions of string compactifications that lead to axion-decay constants \cite{Grimm:2014vva,Carta:2016ynn} and kinetic gauge-couplings in the effective theory \cite{Grimm:2010ks}, this provides new tools for F-theory phenomenology.
\\

\noindent
\textbf{\Large Aknowledgements} \\

\noindent
The original work on which this article is based was done in collaboration with my supervisor Thomas W. Grimm, whom we thank in particular.
We thank the organizers of the ``Workshop of Geometry and Physics'' at Ringberg Castle, Germany, in November 2016 for an inspiring meeting. This article is a summary of a talk that was given on this occasion. In particular we thank Dieter L\"ust for the opportunity to give a talk. Special thanks go to Pierre Corvilain, Irene Valenzuela and Florian Wolf for their support.

\end{document}